\documentclass{elsart}
\usepackage{epsfig}
\usepackage{graphicx}
\usepackage[mathscr]{eucal}
\begin{document}

\begin{frontmatter}

\title{Unmasking the tail of the cosmic ray spectrum}
\author{L. A. Anchordoqui$^a$, M. T. Dova $^b$, T. P.
McCauley$^a$,} \author{S. Reucroft$^a$, and J. D. Swain$^a$}

\address{$^a$Department of Physics, Northeastern University, Boston, MA 02115}
\address{$^b$ Departamento de F\'{\i}sica, UNLP, C.C. 67 (1900) La Plata, Argentina}

\begin{abstract}

A re-examination of the energy cosmic ray spectrum above $10^{20}$
eV is presented. The overall data-base provides evidence, albeit
still statistically limited, that non-nucleon primaries could be
present at the end of the spectrum. In particular, the possible
appearance of superheavy nuclei (seldom discussed in the
literature) is analysed in detail.

\end{abstract}

\end{frontmatter}

\newpage

The origin and nature of cosmic radiation have been a constant
source of mystery and discovery since 1949 \cite{jr}. Most
notably, Greisen, Zatsepin and Kuz'min (GZK) pointed out that
extremely high energy cosmic rays (usually assumed to be nucleons
or nuclei) undergo reactions with the pervasive microwave
background radiation (MBR) yielding a steep drop in their energy
attenuation length \cite{GZK}. Specifically, any proton energy
above 50 EeV is degraded by resonant scattering via $\gamma +p
\rightarrow \Delta \rightarrow p/n + \pi$, and heavy nuclei with
energies above a few tens EeV get attenuated mainly by
photodisintegration off the MBR and intergalactic infrared
background photons (IR). Over the last few years, several giant
air showers have been detected which confirm the arrival of
particles with energies $\geq 100$ EeV, this is, above the GZK
cutoff (see \cite{YD} for a recent survey). Many models have been
proposed as source candidates of such high energy events
\cite{BS}, however, it is not known for certain at the present
time from where the rays originate.

In revealing their origin, the observed anisotropy of these cosmic
rays is one of the most useful features. Very recently, the Fly's
Eye \cite{fe-EeV} and Akeno Giant Air Shower Array (AGASA)
\cite{AGASA-EeV} experiments reported a small but statistically
significant anisotropy  $\mathscr{O}(4\%)$ in the cosmic ray flux
towards the galactic plane at energies around 1 EeV. With
increasing energy the picture looks rather different: although at
$E> 40$ EeV an enhancement of the flux from the Supergalactic
plane was reported \cite{jeremy}, the arrival directions above 100
EeV are best described as isotropic, without imprint of
correlation with the galactic plane or Supergalactic plane
\cite{hillas}. There are two extreme explanations for this puzzle:
i) the bunch of nearby sources follows an isotropic distribution
(which hardly could be the case) ii) One (A few) source(s)
dominates at the highest energies whilst the background fields of
the intergalactic medium strongly modify the particle propagation.
For the latter explanation, it was suggested that a Galactic wind
akin the solar wind could bend all the orbits of the highest
energy cosmic rays towards the Virgo cluster (VC) \cite{a-mt-b-s}.
Actually, if one assumes that these particles are protons, except
for the two highest energy events (the one recorded at AGASA
\cite{haya} and the super-GZK event reported by the Fly's Eye
group \cite{bird}) all trajectories can be traced to within less
than about 20 degrees from Virgo.\footnote{Notice that the highest
energy Yakutsk event was excluded from this sample because of the
great uncertainty on its energy determination. While first
estimates suggested a primary energy around 120 EeV \cite{efimov},
a re-estimation of the number of charged particles at 600 m from
the shower core yields a possible primary energy of 300 EeV
\cite{antonov}.}

At the highest energies, observed extensive air showers seem to be
consistent with nucleon primaries but due to the poor statistics
and large fluctuations from shower to shower an accurate
determination of the particle species is not possible at the
moment. Furthermore, extensive air shower  simulations depend to
some extent on the hadronic interaction event generator which
complicates the  interpretation of data even more \cite{utah}.
Interestingly enough, however, the muon component of the highest
AGASA event agrees with the expectation extrapolated from lower
energies \cite{haya}. Indeed, a population of piled-up protons is
expected at 50 EeV \cite{hs}, and the picture seems quite
consistent. On the other hand, the Fly's Eye event occurs high in
the atmosphere, and, although a primary proton cannot be excluded,
a heavy nucleus more closely fits its shower development
\cite{hvsv}.

It is widely believed that the cosmic ray spectrum beyond the
``crossover energy'' (energy at which the local spectrum becomes
comparable to or less than the cosmological component) could be
associated with the presence of a particularly bright
extragalactic, though relative nearby source, superimposed on a
cosmological diffuse background. In Fig. 1 we show the evolved
energy spectrum of nucleons assuming a cosmologically homogeneus
population of sources -- usually referred to as the universal
hypothesis (UH)-- \cite{syt}, together with a compilation of
recent air shower data \cite{FE-AGASA}. In addition, we show the
modified spectrum for the case of an extended source described by
a Gaussian distribution of width 2 Mpc at a distance of 18.3 Mpc
(see \cite{3k} for details). Assuming that there is no other
significant energy loss mechanism beyond interactions with the MBR
for cosmic rays traversing parts of the cluster, this could be
taken as a very crude model of Virgo. It is important to stress
that for a galactic magnetic field $B_{\rm gal} = 7\mu$G (as in
\cite{a-mt-b-s}) which extends $R_{\rm halo} \sim 1.5$ Mpc in the
galactic halo, the mean flight time of the protons during their
trip through the Milky Way is $\sim 5.05 \times 10^6$ yr
\cite{gustavo}. This means that the bending does not add
substantially to the travel time, and the continuous energy loss
within the straight line approximation is expected to be
reasonable for the problem at hand. From Fig. 1 we realize that
the spectrum of the VC successfully reproduces the AGASA data
above 100 EeV. However, it apparently cannot account for the
super-GZK Fly's Eye event. The interpretation that we give for
this result is that, without specific knowledge of the chemical
composition, the best guess is that at the end of the spectrum two
different types of characters are playing.\footnote{We remark that
AGASA data could be also reproduced if sources of ultra high
energy protons trace the inhomogeneous distribution of luminous
matter in the local present-epoch universe \cite{mt}.}

At this stage, it is interesting to note that the measured density
profile of the highest energy Yakutsk event (excluded in the
analysis of \cite{a-mt-b-s}) shows a huge number of muons.
Remarkably, its arrival direction coincides with the 300 EeV Fly's
Eye event, within angular resolution, possibly indicating a common
origin. If this is the case, the almost completely muonic nature
of this event, recently associated with a dust grain impact
\cite{dg}, could be, perhaps, the signature of a super-heavy
nucleus.

It has been generally thought that $^{56}$Fe is a significant end
product of stellar evolution and higher mass nuclei are rare in
the cosmic radiation. Strictly speaking, the atomic abundances of
middle-weight $(60\leq A < 100)$ and heavy-weight ($A>100$)
elements are approximately 3 and 5 orders of magnitude lower,
respectively, than that of the iron group \cite{BBFH}. The
synthesis of the stable super-heavy nuclides  is classically
ascribed to three different stellar mechanisms referred to as the
s-, r-, and p-processes. The s-process results from the
production of neutrons and their capture by pre-existing seed
nuclei on time scales longer than most $\beta$-decay lifetimes.
There is observational evidence that such a kind of process is
presently at work in a variety of chemically peculiar Red Giants
\cite{smith} and in special objects like FG Sagittae \cite{langer}
or  SN1987A \cite{mazzali}. The abundance of  well developed
nuclides peaks at mass numbers $A=138$ and $A=208$. The
neutron-rich (or r-nuclides) are synthesized when seed nuclei are
subjected to a very intense neutron flux so that $\beta$-decays
near the line of stability are far too slow to compete with the
neutron capture. It has long been thought that appropriate
r-process conditions could be found in the hot ($T \geq 10^{10}K$)
and dense ($\rho \sim 10^{10} - 10^{11}$ g/cm$^{3}$) neutron-rich
(neutronized) material located behind the outgoing shock in a type
II supernova event \cite{delano}. Its abundance distribution peaks
at $A=130$ and $A=195$. The neutron-deficient (or p-nuclides) are
100 - 1000 times less abundant than the corresponding more neutron
rich isobars, while their distribution roughly parallels the s-
and r- nuclides abundance curve. It is quite clear that these
nuclides cannot be made by neutron capture processes. It is
generally believed that they are produced from existing seed
nuclei of the s- or r-type by addition of protons (radiative
proton captures), or by removal of neutrons (neutron
photodisintegration). The explosion of the H-rich envelopes of
type II supernovae has long been held responsible for the
synthesis of these nuclides \cite{BBFH}.

In light of the above, starbursts appear (hopefully) as the
natural sources able to produce relativistic super-heavy nuclei.
These astrophysical environments are supposed to comprise a
considerable population of O and Red Giant stars \cite{sm}, and we
believe the supernovae rate is as high as 0.2 - 0.3 yr$^{-1}$
\cite{m-ua-f}. Of special interest here, the arrival directions of
the Fly's Eye and Yakutsk super-GZK events ($b=9.6^\circ$,
$l=163^\circ$ and $b=3^\circ$, $l=162^\circ$) seem to point
towards the nearby metally-rich galaxy M82 ($b=41^\circ$, $l=
141^\circ$) \cite{es} which has been described as the archetypal
starburst galaxy \cite{fa} and as a prototype of superwind
galaxies \cite{ham}. The joint appearance of the galactic wind and
the galactic magnetic field during particle propagation could
certainly account for the required 37$^\circ$ deflection. In
addition, it was recently suggested that within this type of
galaxies, iron nuclei can be accelerated to extremely high
energies if a two step process is invoked \cite{ngc}. In a first
stage, ions are diffusively accelerated up to a few PeV at single
supernova shock waves in the nuclear region of the galaxy
\cite{cathy}. Since the cosmic ray outflow is convection
dominated, the typical residence time of the nuclei in the
starburst results in $t \sim 1 \times 10^{11}$ s. Thus, the total
path traveled is substantially shorter than the mean free path
(which scales as $A^{-2/3}$) of a super-heavy nucleus (for details
see \cite{ngc}). Those which are able to escape from the central
region without suffering catastrophic interactions could be
eventually re-accelerated to superhigh energies at the terminal
shocks of galactic superwinds generated by the starburst. The
mechanism efficiently improves as the charge number $Z$ of the
particle is increased. For this second step in the acceleration
process, the photon field energy density drops to values of the
order of the cosmic background radiation (we are now far from the
starburst region). The dominant mechanism for energy losses in
the bath of the universal cosmic radiation is the
photodisintegration process \cite{puget}. Notice that the energy
loss rate due to photopair production could be estimated as
$Z^2/A$ times higher than that of a proton with the same Lorentz
factor, and thus could be safely neglected \cite{e+e-}. The
disintegration rate $R$ (in the system of reference where the MBR
is at $2.73K$) of an extremely high energy nucleus with Lorentz
factor $\Gamma$, propagating through an isotropic soft photon
background reads \cite{S},\footnote{Primed quantities refer to the
rest frame of the nucleus.}
\begin{equation}
R = \frac{1}{2\Gamma^2} \int_0^\infty d\epsilon\,
\frac{n(\epsilon)} {\epsilon^2}\,\int_0^{2\Gamma\epsilon}
d\epsilon' \,\epsilon' \, \sigma (\epsilon'),
\end{equation}
where $\sigma$ stands for the total photon absortion cross
section. The density of the soft photon background $n(\epsilon)$
can be  modeled as the sum of: i) the MBR component which follows
a Planckian distribution of temperature $\approx 2.73 K$, ii) the
IR background photons as estimated in \cite{MS}, iii) a black body
spectrum with $T=5000K$ and a dilution factor of $1.2 \times
10^{-15}$ to account for the optical (O) photons. The total photon
absortion cross section is characterized by a broad maximum,
designated as the giant resonance, located at an energy of 12-20
MeV depending on the nucleus under consideration. For the medium
and heavy nuclei, $A\geq 50$, the cross section can be well
represented by a single, or in the case of the deformed nuclei, by
the superposition of two Lorentzian curves of the form
\begin{equation}
\sigma(\epsilon') = \sigma_0 \frac{\epsilon'^2 \,
\Gamma_0^2}{(\epsilon^2_0 - \epsilon'^2)^2 \, + \,\epsilon'^2 \,
\Gamma_0^2}.
\end{equation}
In order to make some estimates, hereafter we refer our
calculations to a gold nucleus (the resonance parameters are
listed in table I \cite{FW}). In Fig. 2 we show the $^{197}$Au
photodisintegration rate due to interactions with the starlight
and relic photons. At the highest energies, the energy losses are
dominated by collisions with the tail of $2.73 K$ Planckian
spectrum. It is straightforward to show that a  superheavy nucleus
of a few hundred EeV emitted by M82 can traverse almost unscathed
through the primeval radiation to produce an extensive air shower
after interaction with the earth atmosphere.

\begin{table}
\caption{Giant dipole resonance parameters}
\begin{tabular}{ccc}
\hline\hline $\epsilon_0$ [MeV] & $\sigma_0$ [mb] & $\Gamma_0$
[MeV]
\\ \hline 13.15 & 255 & 2.9 \\ 13.90 & 365 & 4.0 \\
\hline \hline
\end{tabular}
\end{table}

Additional support for the superheavy nucleus hypothesis comes
from the CASA-MIA experiment \cite{casamia} (See in particular
Fig. 9). The collected cosmic ray data between $10^{14}$ -
$10^{16}$ eV tends to favor a supernova shock wave acceleration
scenario. The average mass increases with energy, becoming heavier
above $10^{15}$ eV. At the maximum energy the results are
consistent at 1$\sigma$ level with nuclei heavier than iron.
However, ``lore'' has settled down some comparisons of the
admittedly limited ultra high energy cosmic ray sample against
hadronic-interaction-event generators which predicts the arrival
of particle species heavier than iron \cite{casamia,nagano-etal}.
We would like to stress that since simulations are used to
interpret data, and then the data is used to modify the
simulation, one has to be very careful and as we have shown, it is
by no means clear that superheavy nuclei could not be present at
the end of the spectrum.

The energy spectrum of nearby (around 3Mpc) nuclear sources was
discussed elsewhere \cite{depression}. The analysis showed that
particles tend to pile up between 240 - 270 EeV. This bump-like
feature is followed by a simultaneous drop in the cosmic ray flux
of the preceding bins of energy, changing the relative detection
probabilities. As a consequence particles in the pile-up are 50\%
more probable than those at lower energies.

In summary, the recently reported AGASA data can be successfully
reproduced by a power law spectrum of nucleons hailing from the VC
superimposed on a cosmological diffuse background. The Fly's Eye
observations may also fit this scenario, albeit with large errors.
One might also consider less likely astrophysical sources. In
particular, our analysis seems to indicate that the next-door
galaxy M82 could be responsible for some events at the end of the
CR spectrum. This has also been suggested elsewhere
\cite{es,ngc,depression}. At least some of these super-GZK events
could be due to heavy, and even superheavy nuclei. Clearly, more
data is needed before this hypothesis can be verified. In this
regard, the coming avalanche of high quality cosmic ray
observations at the Southern Auger Observatory \cite{auger} will
provide new insights to the ideas discussed in this letter.

{\em Note added}: After we finished this work, it was argued that
the Galactic wind model assumed in Ref. \cite{a-mt-b-s} is alone
responsible for the focusing of positive particles towards the
North galactic pole. Therefore the apparent clustering of the
back-traced CR cannot be interpreted as evidence for a point
source, this point source identified  as  M87 \cite{b-l}. It
should be pointed out that the main input parameters for the
determination of the CR-spectrum in Fig. 1 are the spectral index
of the source, and the propagation distance of the nucleons in the
extragalacic medium. The Galactic wind model is just used to
collect all the traces in only one single direction in the sky.
Therefore the discussion presented in this letter strongly
supports older suspicions regarding M87, like the model proposed
in Ref. \cite{a-mt-b-s}.

\hfill

In closing, we wish to thank Gustavo Medina Tanco for a fruitful
discussion. The research of LAA was supported by CONICET. MTD was
supported by CONICET and Fundaci\'on Antorchas. TPM-SR-JDS were
supported by the National Science Foundation.

\begin{figure}
\begin{center}
\epsfig{file=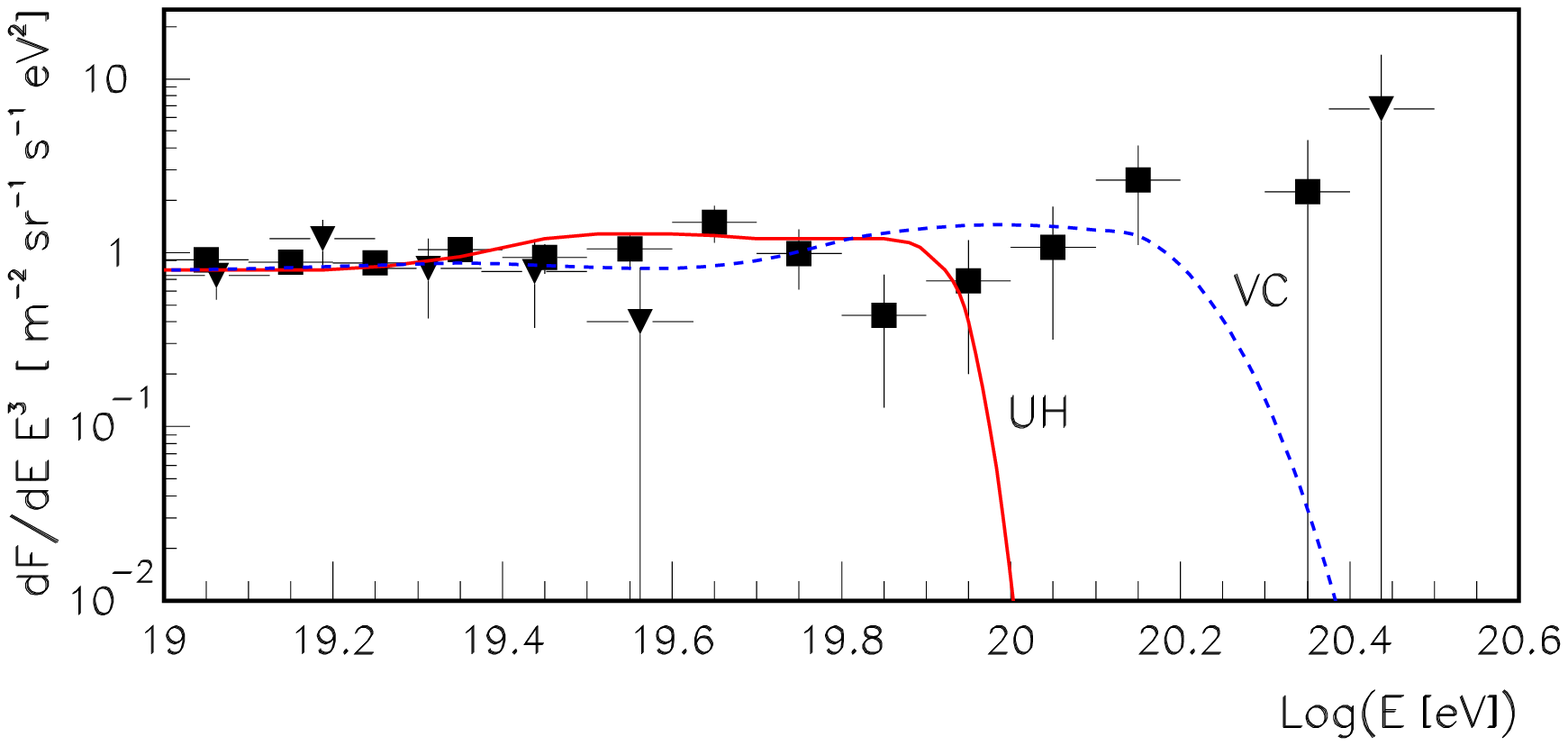,width=14.5cm,clip=} \caption{The cosmic ray
flux spectrum derived from AGASA (square) and Fly's Eye (triangle)
experiments shown with the shape of the universal hypothesis (UH)
spectrum (spectral index $\gamma = 3.27$ [20]). We also show the
expected flux of ultra high energy nucleons from the Virgo cluster
(VC).}
\end{center}
\end{figure}

\newpage

\begin{figure}
\begin{center}
\epsfig{file=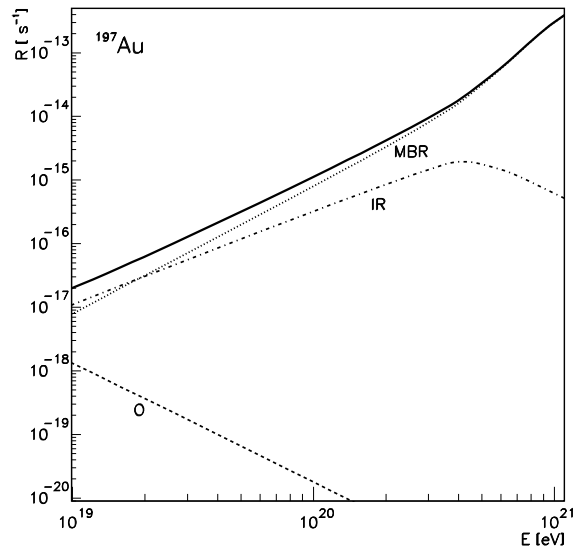,width=7.5cm,clip=} \caption{Fractional
energy loss for $^{197}$Au photodisintegration on MBR, IR,
O, as well as the total (solid line).}
\end{center}
\end{figure}

\end{document}